\documentclass[12pt]{article}

\usepackage{amsmath, euscript, amssymb, amsfonts, latexsym}

\usepackage{mathrsfs}

\newcommand{\mS}{{\mathscr{S}}}
\newcommand{\mW}{{\mathscr{W}}}

\newcommand{\mM}{{\mathscr{M}}}
\newcommand{\mH}{{\mathscr{H}}}

\newcommand{\oR}{{\mathbb R}}
\newcommand{\oV}{{\mathbb V}}

\newcommand{\supp}{\mathop{\mathrm{supp}}\nolimits}

\renewcommand{\Re}{\mathop{\mathrm{Re}}\nolimits}
\renewcommand{\Im}{\mathop{\mathrm{Im}}\nolimits}
\newcommand{\eqdef}{\stackrel{\text{\tiny def}}{=}}

\begin{document}

\bigskip

\hfill FIAN-TD/2008-01

\baselineskip=20pt

\vspace{2cm}

\begin{center}

{\Large\bf Failure of microcausality in noncommutative field
theories}

\vspace{1cm}

{\bf M.~A.~Soloviev}\footnote{E-mail: soloviev@lpi.ru}

\vspace{0.5cm}

 \centerline{\sl P.~N.~Lebedev Physical Institute}
 \centerline{\sl Russian Academy of Sciences}
 \centerline{\sl  Leninsky Prospect 53, Moscow 119991, Russia}

\vskip 3em

\end{center}

\begin{abstract}
We revisit the question of microcausality violations in  quantum
field theory on noncommutative spacetime, taking $\mathcal
O(x)=:\phi\star\phi:(x)$ as a sample observable. Using methods of
the theory of distributions, we precisely describe the support
properties of the commutator $[\mathcal O(x),\mathcal O(y)]$ and
prove that, in the case of space-space noncommutativity, it does
not vanish at spacelike separation in the non\-commuting
directions. However, the matrix elements of this commutator
exhibit a rapid falloff along an arbitrary spacelike direction
irrespective of the type of non\-commutativity. We also consider
the star commutator for this observable and show that it fails to
vanish even at spacelike separation in the commuting directions
and completely  violates causality. We conclude with a brief
discussion about the modified Wightman functions which are vacuum
expectation values of the star products of fields at different
spacetime points.
\end{abstract}

\vskip 2em

%\newpage
\section{\large Introduction}

In recent years, considerable attention has been given to the
construction of quantum field theories (QFTs) on noncommutative
spacetimes (see, e.g.,~\cite{Sz} for a review). The question of
causality is a basic one in the development of the corresponding
conceptual framework. A noncommutative deformation of the
$d$-dimensional spacetime  is formally defined by replacing the
coordinates $x^\mu$ of $\oR^d$ by  operators $\hat x^\mu$
satisfying the commutation relations
\begin{equation}
[\hat x^\mu, \hat x^\nu]=i\theta^{\mu\nu},
 \label{1.1}
\end{equation}
where $\theta^{\mu\nu}$ is a real antisymmetric  $d\times
d$-matrix, constant in the simplest case. However, this
deformation can be combined with the basic principles of quantum
theory in a variety of fashions. In particular, the issue of
causality cannot be discussed in isolation from that of the
implementation of spacetime symmetries. The relations~\eqref{1.1}
are not covariant under the Lorentz transformations, and
noncommutative QFT is usually treated as a specific form of field
theory with a nonlocal interaction breaking the Lorentz symmetry
to a subgroup. In the Lagrangian formalism, the theory is defined
by replacing the ordinary product of fields in the interaction
terms of the actions with the Moyal $\star$-product given by
\begin{equation}
(\phi_1\star
\phi_2)(x)=\phi_1(x)\exp\left(\frac{i}{2}\,\overleftarrow{\partial_\mu}\,
\theta^{\mu\nu}\,\overrightarrow{\partial_\nu}\right)\phi_2(x).
 \label{1.2}
\end{equation}
 The star product commutation
relation $x^\mu \star x^\nu-x^\nu \star x^\mu=i\theta^{\mu\nu}$ is
identified with~\eqref{1.1} via the Weyl correspondence between
operators and their symbols. There is an essential distinction
between the cases of a space-space and time-space
noncommutativity. If the time coordinate is involved in
noncommutativity, then a string theoretical interpretation of the
field theory comes up against the problem of
nonunitarity~\cite{GM} and  inconsistency with the conventional
Hamiltonian evolution~\cite{SST}. A consistent  Hamiltonian
framework for the scalar field theory with time-space
noncommutativity has been proposed in~\cite{BDFP}. The definition
given there leads to an perturbatively unitary S-matrix and is
interesting by itself, even though its relationship with string
theory is unclear. Field theories with only space noncommutativity
(that is $\theta^{0\nu}=0$) avoid the problems with unitarity, and
models of this form  attract the most notice because they describe
a low energy limit of string theory in certain backgrounds.
However, its causal structure is  different from that of the
standard QFT because the light cone is changed to a light wedge
respecting the residual Lorentz symmetry~\cite{Alv,LS,CFI}. The
main object of the present paper is to analyze rigorously this
modification of the causal structure.

At present, much consideration is  being given to quantization of
noncommutative theories with the use of a ``twisted'' version of
the Poincar\'e covariance. These efforts are aimed at restoring
the spacetime symmetries broken by noncommutati\-vity and
developing a covariant formulation  even though the matrix
$\theta^{\mu\nu}$ in~\eqref{1.1}, \eqref{1.2} is constant. Within
this setting, the issues of locality and causality were discussed,
e.g., in~\cite{FW,BPQV,RSz}, but up to now there  is no consensus
regarding  the implementation of the twisted covariance in QFT and
its physical consequences. Another way of looking at
noncommutative spacetime was proposed in~\cite{GL}, where an
infinite family of fields labelled by different noncommutativity
parameters was considered and their relative localization
properties were investigated.

In~\cite{Ch1},  locality and causality violations caused by
non\-commutativity  were illustrated by a star product analogue of
the normal  ordered square $:\phi^2:$ of a free scalar field
$\phi$. Specifically, Chaichian {\it et al} considered $\mathcal
O(x)={:\phi\star\phi:(x)}$ as a sample observable and found that
the matrix element
\begin{equation}
\langle 0|[\mathcal O(x)), \mathcal
O(y)]\Bigr|_{x^0=y^0}|p_1,p_2\rangle
 \label{1.3}
\end{equation}
is nonzero only when $\theta^{0\nu}\ne 0$. More recently,
Greenberg~\cite{G} considered the  commutator $[\mathcal O(x),
\partial_\nu\mathcal O(y)]$ with the derivatives of $\mathcal O$
and has shown that it fails to vanish at equal times even in the
case in which $\theta^{0\nu}= 0$. As stated in~\cite{G}, this
result holds generally when there are time derivatives in the
observables. A similar conclusion was reached in~\cite{HJ}, where
also a commutator involving time derivatives was treated, but with
the use of a generalization of the Bogoliubov-Shirkov causality
criterion.

In this paper,   we  analyze  the commutator $[\mathcal O(x),
\mathcal O(y)]$ more closely, using the techniques of the theory
of distributions, which allows  describing  its support properties
completely. We first consider the case of space-space
noncommutativity, taking  for definiteness  $d=4$ and
$\theta^{12}=-\theta^{21}\ne 0$, with the other values of the
$\theta$-matrix equal to zero. In Sec.~2, we show that then the
commutator vanishes in the spacelike wedge
$|x^0-y^0|\leq|x^3-y^3|$. In Sec.~3, we prove that $[\mathcal
O(x), \mathcal O(y)]\ne 0$ everywhere outside this wedge. This
result demonstrates that the space-space noncommutativity violates
the usual  $SO(1,3)$ microcausality even if there are no time
derivatives in the observables. In Sec.~4, we show that
nevertheless the matrix elements of the commutator decrease
rapidly in the whole   cone $(x-y)^2\leq0$ and behave like
$\exp(-|x-y|^2/|\theta|)$ at large spacelike separation. This is
true without regard to the type of noncommutativity, in both
space-space and time-space cases, and manifests itself after
averaging the observable $\mathcal O(x)$ with  sufficiently smooth
and rapidly decreasing test functions. The best suitable  class of
test functions has been found and investigated  in~\cite{S1,S2}. A
slightly different class was independently proposed in~\cite{Ch2}.
In Sec.~5, we examine the modified commutator  $[\mathcal O(x),
\mathcal O(y)]_{\star}=\mathcal O(x)\star\mathcal O(y)-\mathcal
O(y)\star\mathcal O(x)$,  where    $\star$  denotes now a star
multiplication of field operators at different spacetime points.
Such a modification was also discussed in the literature. We prove
that, contrary to expectations,  the star commutator fails to
vanish even in the spacelike wedge and completely violates
causality. Our study shows in particular, that the seemingly
natural definition of the star product of fields at different
spacetime points, as an operation dual to the corresponding
operation on test functions, brings the causality principle and
the spectral condition into conflict. Section~6 contains
concluding remarks.

\section{\large A light wedge instead of the light cone}

Let $\phi$ be a free neutral scalar field of mass $m$ on a
spacetime of $d$ dimensions and let $:\phi^2:(x)= \lim_{x_1, x_2
\rightarrow x}:\phi(x_1)\phi(x_2):$. By the Wick theorem for
normal ordered products, it follows that
\begin{equation}
 \langle 0|:\phi^2\!:(x):\phi^2\!:(y):\phi(z_1)\phi(z_2):|0\rangle=
 4w(x-y)w(x-z_1)w(y-z_2)+ (z_1\leftrightarrow z_2),
 \label{2.1}
\end{equation}
where $w$ is the two-point function of $\phi$, i.e.,
\begin{equation}
w(x-y)= \langle
0|\phi(x)\phi(y)|0\rangle=\frac{1}{(2\pi)^{d-1}}\int dk\,
\vartheta(k^0)\delta(k^2-m^2)e^{-ik\cdot (x-y)}.
 \label{2.2}
\end{equation}
As a consequence, we have
\begin{equation}
\langle
0|[:\phi^2\!:(x),:\phi^2\!:(y)]\,:\phi(z_1)\phi(z_2):|0\rangle=
 4i\Delta(x-y)w(x-z_1)w(y-z_2)+ (z_1\leftrightarrow z_2),
 \label{2.3}
\end{equation}
where $\Delta(x-y)=\frac{1}{i(2\pi)^{d-1}}\int dk\,
\epsilon(k^0)\delta(k^2-m^2)e^{-ik\cdot(x-y)}$ is the Pauli-Jordan
function. Let us  now consider the normal ordered  expression
\begin{multline}
\mathcal O(x)=:\phi\star\phi:(x) = \lim_{x_1, x_2 \rightarrow x}
 :\phi(x_1)\exp\left(\frac{i}{2}\,\overleftarrow{\partial_\mu}\,
\theta^{\mu\nu}\,\overrightarrow{\partial_\nu}\right)\phi(x_2)\!:\,=\\
:\phi^2:(x)+\sum_{n=1}^\infty\left(\frac{i}{2}
\right)^n\frac{1}{n!}\,\theta^{\mu_1\nu_1}\dots
\theta^{\mu_n\nu_n}:\partial_{\mu_1}\dots
\partial_{\mu_n}\phi(x)\,\partial_{\nu_1}\dots\partial_{\nu_n}\phi(x):.
 \label{2.4}
\end{multline}
Every term in the expansion~\eqref{2.4} is well defined as a Wick
monomial in derivatives of $\phi$, see~\cite{SW} or~\cite{BLOT}.
The technique developed in~\cite{SS} allows us to define
rigorously their sum as an operator-valued  generalized function
acting in the Hilbert space of $\phi$, but we will not dwell on
this point and now restrict our consideration to the vacuum
expectation value
\begin{equation}
\mW(x,y;z_1,z_2)=\langle 0|\,\mathcal O(x)\mathcal
O(y):\phi(z_1)\phi(z_2)\!:|0\rangle,
 \label{2.5}
 \end{equation}
 which is an analogue of~\eqref{2.1}. Applying the Wick theorem
 again and using the formula
\begin{equation}
 e^{ik\cdot x}\star e^{ip\cdot x}=e^{-i[k,p]}e^{i(k+p)\cdot
 x},
 \label{2.6}
\end{equation}
where
$$
 [k,p]\eqdef
 (1/2)k_\mu\theta^{\mu\nu}p_\nu,
$$
one can readily see that
\begin{multline}
 \mW(x,y;z_1,z_2)=
 4\!\int dkdp_1dp_2\,\tilde w(k)
 e^{-ik\cdot (x-y)-ip_1\cdot (x-z_1)-ip_2\cdot (y-z_2)}\\
 \times\prod_{i=1,2} \tilde w(p_i)\cos[k,p_i] +
(z_1\leftrightarrow z_2),
 \label{2.7}
 \end{multline}
 where $\tilde w(k)=\left(\mathcal Fw\right)(k)=\int d\xi\, e^{ik\cdot\xi}w(\xi)$.
 More explicitly, the Fourier transform of~\eqref{2.5} has the
 form
 \begin{multline}
 \tilde{\mW}(k_1,k_2;p_1,p_2)=
4(2\pi)^{d+3}\delta(k_1+k_2+p_1+p_2)\vartheta(k^0_1+p^0_1)\delta((k_1+p_1)^2-m^2)\\
\times\prod_{i=1,2}
\vartheta(-p_i^0)\delta(p_i^2-m^2)\cos[k_i,p_i]
+(p_1\leftrightarrow p_2),
 \label{2.8}
\end{multline}
where $k_1$, $k_2$ and  $p_1$, $p_2$ are the momentum-space
variables conjugate to  $x$, $y$ and $z_1$, $z_2$, respectively.
The function
\begin{equation}
\mu=\cos[k_1, p_1]\cos[k_2,p_2]
 \label{2.9}
\end{equation}
is a multiplier of the Schwartz space $\mS(\oR^{4d})$ and hence
the expression on right-hand side of~\eqref{2.8} and the vacuum
expectation value~\eqref{2.5} are well defined as tempered
distributions.

 From Eq.\eqref{2.7}, it follows that
\begin{multline}
 \langle 0|[\mathcal O(x),\mathcal
O(y)]\,:\phi(z_1)\phi(z_2)\!:|0\rangle\\
= 4i\int\!\! dkdp_1dp_2\tilde \Delta (k)
   e^{-ik\cdot (x-y)-ip_1\cdot (x-z_1)-ip_2\cdot (y-z_2)}
   \prod_{i=1,2} \tilde w(p)\cos[k,p_i] +
(z_1\leftrightarrow z_2),
 \label{2.10}
\end{multline}
which agrees with formulas for the matrix element $\langle
0|[\mathcal O(x),\mathcal O(y)]|p_1, p_2\rangle$ in~\cite{Ch1} and
\cite{G}. The Fourier transform of distribution~\eqref{2.10} is
obtainable by multiplying that of~\eqref{2.3} by the
multiplier~\eqref{2.9}. The distribution~\eqref{2.3} is zero
everywhere in the cone $(x-y)^2<0$, but this is not to say that
the distribution~\eqref{2.10} obeys microcausality. Let us turn to
the case of space-space noncommutativity, assuming that
$\theta^{12}=-\theta^{21}=\theta\ne 0$ and the  other elements of
the matrix  $\theta^{\mu\nu}$ are equal to zero. It is easily seen
that then the distribution~\eqref{2.10} satisfies a weakened
version of microcausality and vanishes in the wedge defined by
\begin{equation}
 |x^0-y^0|<|x^3-y^3|.
 \label{2.11}
\end{equation}
In fact, the Fourier transformation converts multiplication into
convolution\footnote{To be more precise, we use the relations $(u,
\tilde g)=(\tilde u, g)$ and $\widetilde{\mu
g}=(2\pi)^{-d}\tilde\mu\,*\tilde g$, which hold for any
 $g\in \mS(\oR^d)$,  $u\in \mS'(\oR^d)$ and for each multiplier $\mu$.}
  and hence  the value of distribution~\eqref{2.10} at a test
  function $f$ coincides with the value of~\eqref{2.3} at the test function
$(2\pi)^{-4d}\,\tilde\mu*f$. Under our assumptions about the
$\theta$-matrix, the multiplier~\eqref{2.9} does not dependent on
the variables conjugate to $x^0,y^0,x^3,y^3$, and its Fourier
transform $\tilde\mu$ is the tensor product of
$\delta(x^0)\delta(y^0)\delta(x^3)\delta(y^3)$ and a distribution
in  the other variables. Therefore, if  $\supp f$ is contained in
the wedge~\eqref{2.11}, then  $\supp(\tilde\mu*f)$ also lies in
this wedge and does not intersect the support of
distribution~\eqref{2.3}. It follows that the
distribution~\eqref{2.10} vanishes for such test functions.

\section{\large Violations of microcausality}

Now we intend to show that in the case of space-space
noncommutativity, the commutator  $[\mathcal O(x),\mathcal O(y)]$
does not vanish outside the wedge~\eqref{2.11} and hence the
observable $\mathcal O(x)$ defined by~\eqref{2.4}  does not
satisfy the standard microcausality condition.

\medskip

 {\bf Theorem 1.} {\it  Let $d=4$ and let
$\theta^{12}=-\theta^{21}=\theta\ne 0$, with the other elements of
the  matrix $\theta^{\mu\nu}$ equal to zero. Suppose that points
$\bar x,\bar y \in \oR^4$ satisfy the inequalities  $(\bar x-\bar
y)^2<0$ and $|\bar x^0-\bar y^0|>|\bar x^3-\bar y^3|$. Then there
is a state $\Phi$ such that $(\bar x,\bar y)$ belongs to the
support of
\begin{equation}
\mM_\Phi(x,y)\eqdef\langle 0|[\mathcal O(x),\mathcal
O(y)]|\Phi\rangle.
 \label{3.1}
\end{equation}

\medskip

 Proof.} We take a state of the form
\begin{equation}
 |\Phi\rangle=\int\!dz_1dz_2\,:\phi(z_1)\phi(z_2)\!: h(z_1)h(z_2)|0\rangle=
 \phi^-(h)\phi^-(h)|0\rangle,
  \label{3.2}
\end{equation}
where $h\in \mS(\oR^4)$. Then $\mM_\Phi$ is clearly a tempered
distribution  and by~\eqref{2.10} we have
\begin{multline}
(\mM_\Phi, f\otimes g)=\langle 0|[\mathcal O(f),\mathcal
O(g)]|\Phi\rangle=\\
8i\int dp_1dp_2 \prod_{i=1,2}\tilde w(p_i) \tilde h(p_i)\int dk\,
\tilde \Delta(k) \cos[k,p_1]\cos[k,p_2]\tilde f(-k-p_1)\tilde
g(k-p_2)
 \label{3.3}
\end{multline}
for any test functions $f,g\in \mS(\oR^4)$. This order of
integration is permissible by the Fubini theorem because
integrating over $k^0$, $p^0_1$, $p^0_2$ gives an integrable
function  on $\oR^9$. The function
 $$
 \psi_{p_1,p_2}(k)=\tilde f(-k-p_1)\tilde
g(k-p_2)
$$
belongs to the space $\mS(\oR^4)$ for any $p_1,p_2\in \oR^4$, and
the function
$$
\mu_{p_1,p_2}(k)=\cos[k,p_1]\cos[k,p_2]
$$
is a multiplier of  $\mS(\oR^4)$. Therefore the integral over $k$
in~\eqref{3.3} can be written as
\begin{equation}
(\tilde\Delta, \mu_{p_1,p_2}\cdot
\psi_{p_1,p_2})=\frac{1}{(2\pi)^4}(\Delta,\tilde\mu_{p_1,p_2}*\tilde
\psi_{p_1,p_2})
  \label{3.4}
\end{equation}
  Let $\Theta$ be the linear map defined by $(\Theta
p)^\mu=\frac{1}{2}\theta^{\mu\nu}p_\nu$. Then
\begin{multline}
\tilde\mu_{p_1,p_2}(\xi)=\int dk\,
e^{ik\cdot\xi}\mu_{p_1,p_2}(k)\\=
\frac{(2\pi)^4}{4}[\delta(\xi-\Theta(p_1+p_2))+\delta(\xi+\Theta(p_1+p_2))
+\delta(\xi-\Theta(p_1-p_2))+\delta(\xi+\Theta(p_1-p_2))]. \notag
 \end{multline}
Furthermore, we have
\begin{multline}
\tilde \psi_{p_1,p_2}(\xi)=\int dk\,
e^{ik\cdot\xi}\psi_{p_1,p_2}(k)=\int dkdxdy\,
e^{ik\cdot\xi+i(-k-p_1)\cdot x+i(k-p_2)\cdot y}f(x)g(y)\\=
(2\pi)^4\int dxdy\,e^{-ip_1\cdot x-ip_2\cdot
y}\delta(\xi-x+y)f(x)g(y)= (2\pi)^4
e^{-i(p_1-p_2)\cdot\xi/2}\varphi_{p_1,p_2}(\xi),
 \notag
 \end{multline}
where
\begin{equation}
\varphi_{p_1,p_2}(\xi)=\int dX\,e^{-i(p_1+p_2)\cdot
X}f(X+\xi/2)g(X-\xi/2).
 \label{3.5}
 \end{equation}
In what follows, we set $\bar y=-\bar x$ and $\bar x^3=\bar
y^3=0$. This does not result in any loss of generality because the
distribution~\eqref{2.10} is invariant under translations and
under boosts  in the $x^3$-direction. If  $\supp f$ is contained
in the $\varepsilon$-neighborhood of $\bar x$ and $\supp g$ is
contained in the  $\varepsilon$-neighborhood of $-\bar x$, then
only points $X$ with $\|X\|\le \varepsilon$ contribute in the
integral in~\eqref{3.5} and  the functions $\varphi_{p_1,p_2}$,
$\tilde \psi_{p_1,p_2}$ have support in the
$2\varepsilon$-neighborhood of the point $2\bar x$. We also note
that the operation consisting in convolution with
 $\tilde\mu_{p_1,p_2}$ displaces $\supp\tilde \psi_{p_1,p_2}(\xi)$ by
 the vectors $\pm\Theta(p_1\pm p_2)$. Now we specify the choice of
 $h$, setting
$$
\bar p^1=2\bar x^2/\theta,\quad \bar p^2=-2\bar x^1/\theta, \quad
\bar p^3=0, \quad \bar p^0=\sqrt{m^2+(\bar p^1)^2+ (\bar p^2)^2},
$$
so that $\Theta\bar p=(0, \bar x^1, \bar x^2,0)$. We take $\tilde
h(p)$ to be a nonnegative function supported in a neighborhood $U$
of $\bar p$ and such that $\tilde h(\bar p)> 0$. We choose $U$ so
small that the set of points $2\bar x\pm \Theta(p_1-p_2)$, where
$p_1$ and $p_2$ run through $U$, is separated from the cone
$\bar\oV=\{\xi\in\oR^4 \colon \xi^2\ge 0\}$ by a positive
distance. Then for any $p_1,p_2\in U$,  one of the four functions
obtained from $\tilde \psi_{p_1,p_2}$ by  convolution with
 $\tilde\mu_{p_1,p_2}$ has support in a neighborhood of
$\bar \xi=(2\bar x^0, 0,0,0)$, whereas the other three of them are
supported in the spacelike region and  do not contribute in the
right-hand side of~\eqref{3.4} if $\varepsilon$ is  small enough.
Inside the cone $\bar \oV$, the distribution $\Delta(\xi)$ is a
regular function and  we have the well-known representation
$$
\Delta(\xi)=\frac{m}{4\pi\sqrt{\xi^2}}\epsilon(\xi^0)J_1(m\sqrt{\xi^2}),\qquad
\xi\in \oV.
$$
We first assume that  $J_1(2m|\bar x^0|)\ne0$ and  impose two
additional restrictions on $\supp \tilde h$. Namely, we choose $U$
so small that $J_1(m\sqrt{\xi^2})$ has a constant sign on the set
\begin{equation}
\{\xi\in \oR^4\colon \xi=2\bar x- \Theta(p_1+p_2),\quad p_1,p_2\in
U\}
 \label{3.6}
 \end{equation}
and furthermore the inequality
\begin{equation}
|(p_1-p_2)\cdot\bar x|< \pi/4
 \label{3.7}
 \end{equation}
holds for all $p_1,p_2\in U$. We put $f(x) =g(-x)$ and assume that
$f(x)\geq 0$, $f(\bar x)\ne 0$. Then the function
$\varphi_{p_1,p_2}(\xi)$ is real because the product
$f(X+\xi/2)f(-X+\xi/2)$ is invariant under the reflection $X\to
-X$. If $\varepsilon$ is sufficiently small, then
$\varphi_{p_1,p_2}(\xi)$ is nontrivial and nonnegative for all
$p_1,p_2\in U$. From~\eqref{3.7}, it follows that $\Re\tilde
\psi_{p_1,p_2}(\xi)$ also has these properties. The support of the
shifted function $\psi_{p_1,p_2}(\xi-\Theta(p_1+p_2))$ lies in the
$2\varepsilon$-neighborhood of the set~\eqref{3.6} and, if
$\varepsilon$ is sufficiently small, then $J_1(m\sqrt{\xi^2})$ has
a constant sign on this support. Therefore the expression
$\Re(\Delta,\tilde\mu_{p_1,p_2}*\tilde \psi_{p_1,p_2})$ has a
constant sign for all $p_1,p_2\in \supp \tilde h$. We conclude
that for arbitrarily small neighborhoods of the points $\bar x$
and $\bar y$, there exist test functions $f$ and $g$ supported in
these neighborhoods and such that  $\langle 0|[\mathcal
O(f),\mathcal O(g)]|\Phi\rangle\ne0$. This amounts to saying that
$(\bar x, \bar y)$ belongs to $\supp \mM_\Phi$. If $J_1(2m|\bar
x^0|)=0$ and $U$ is small enough, then the function
$J_1(m\sqrt{\xi^2})$ has a constant sign on the set~\eqref{3.6}
except for $\xi=2\bar x^0$, and we arrive at the same conclusion
with a different choice of $f$. Namely, we can take $f$ to be a
nonnegative function  supported in the
$\varepsilon/2$-neighborhood of the point $(\bar x^0\pm
\varepsilon/2, \bar x^1, \bar x^2, \bar x^3)$, where the minus
sign corresponds to positive  $\bar x^0$ and the plus sign
corresponds to negative $\bar x^0$. This completes the proof of
Theorem~1.

\medskip

{\large Remark~1.} It is worth noting that this theorem also holds
for $\bar x^0=\bar y^0$, $\bar x^3=\bar y^3$, $(\bar x-\bar
y)^2<0$. In other words, the support of the commutator under study
contains even the equal-time points which lie outside the
wedge~\eqref{2.11}. The proof proceeds along the same lines, but
in this case $f$ should be chosen so that its support is contained
in the $\varepsilon/2$-neighborhood of the point $(\varepsilon/2,
\bar x^1,\bar x^2, 0)$.

\medskip

{\large Remark~2.} Theorem~1 implies, in particular, that the
power series expansion of the distribution~\eqref{2.10} in
$\theta$ does not converge in the topology of the space $\mS'$ of
tempered distributions. In fact, every term of this expansion is
obtainable from~\eqref{2.3} by applying a finite-order
differential operator and hence is  zero everywhere in the region
$(x-y)^2<0$. If the expansion were convergent in $\mS'$, its limit
should also vanish in this region. A weaker topology, in which the
expansion in powers of $\theta$ converges, is indicated
in~\cite{S2}.

\section{\large $\theta$-locality}

We now show that the distribution $\mM_\Phi(x,y)$, if smoothed
properly, has a rapid decrease in the whole cone $(x-y)^2<0$ for
all $\Phi$ ranging a dense set in the subspace of two-particle
states. More precisely, it behaves like $\exp(-|x-y|^2/|\theta|)$
at large spacelike separation of the arguments.\footnote{Here and
in the sequel we use the notation $|\theta|=
\sum_{\mu<\nu}|\theta^{\mu\nu}|$.} This is true irrespectively of
the form of the matrix $\theta^{\mu\nu}$ and, in particular, for
both space-space and time-space noncommutativity.

A simple and well-known way of describing the behavior of a
distribution at infinity is by considering its convolution with
test functions decreasing sufficiently fast. In order to reveal
the indicated decrease of $\mM_\Phi$, it is natural to use test
functions satisfying the inequalities
\begin{equation}
 |\partial^\kappa f(x)|\le C_\kappa e^{-|x/A|^2},
 \label{4.1}
\end{equation}
where $A$ is small in comparison to $\sqrt{|\theta|}$. In our
case, however, the  test functions should also be  sufficiently
smooth, as it is argued in~\cite{S1,S2}. The
distribution~\eqref{2.10} is obtained from the
distribution~\eqref{2.3} by applying the infinite-order
differential operator
\begin{equation}
D_\theta=\cos\left(\frac{1}{2}\partial_x\theta
   \partial_{z_1}\right)\cos\left(\frac{1}{2}\partial_y\theta
   \partial_{z_2}\right), \qquad \partial_x\theta\partial_z\eqdef
   \frac{\partial}{\partial
x^\mu}\theta^{\mu\nu}\frac{\partial}{\partial z^\nu} .
 \label{4.2}
\end{equation}
The function space defined by~\eqref{4.1} is not invariant under
the action of the basic Moyal operator defining the
$\star$-product and  under the action of $D_\theta$. In other
words, these operators spoil in general the behavior of its
elements at infinity. Theorem 2 of~\cite{S1} characterizes those
subspaces of the Schwartz space that are invariant under the Moyal
operator and shows that the smoothness properties of their
elements should be matched with the decrease properties to ensure
this invariance. A special role is played by the space denoted
in~\cite{GS2} by $S^{1/2}_{1/2}$, which consists of the infinitely
differentiable functions satisfying
\begin{equation}
 |\partial^\kappa f(x)|\le C B^{|\kappa|}
\kappa^{\kappa/2}e^{-|x/A|^2},
 \label{4.3}
\end{equation}
where $C$, $B$, $A$ are positive constants depending on $f$ and
the usual multi-index notation is used. This space is the union of
the Banach spaces $S^{1/2,B}_{1/2, A}$ with the norms
\begin{equation}
 \|f\|_{A,B}=\sup_{\kappa, x}e^{|x/A|^2}\frac{|\partial^\kappa f(x)|}{B^{|\kappa|}
\kappa^{\kappa/2}},
 \label{4.4}
\end{equation}
and a sequence $f_n$ is said to be convergent to zero in
$S^{1/2}_{1/2}$ if there  are $A$ and $B$ such that  $f_n\in
S^{1/2,B}_{1/2, A}$ and $\|f_n\|_{A,B}\to 0$ as $n\to\infty$. The
space $S^{1/2}_{1/2}$ is invariant under both the Fourier operator
and the Moyal operator and these operators are continuous in its
topology.

\medskip

{\bf Theorem 2.} {\it Let $\phi$ be a free scalar field on $\oR^d$
and let $\mathcal O(x)=:\phi\star\phi:(x)$, with the
$\star$-product defined by an arbitrary real antisymmetric matrix
$\theta^{\mu\nu}$. Let $f,g, h_1, h_2\in S^{1/2,B}_{1/2, A}$,
where $A>0$ and $0<B<1/\sqrt{e|\theta|}$. Suppose that  $a$  is a
spacelike vector in $\oR^d$ separated from the cone $\bar \oV$ by
an angular distance $\gamma=\inf_{\xi^2\geq 0}|\xi-a/|a||$. Then
the matrix element
\begin{equation}
(\mM_\Phi,f_a\otimes g_{-a})=\langle 0|[\mathcal O(f_a),\mathcal
O(g_{-a})]|\Phi\rangle,
 \label{4.5}
\end{equation}
where  $f_a(x)=f(x-a)$ and
$|\Phi\rangle=\phi^-(h_1)\phi^-(h_2)|0\rangle$, satisfies the
estimate
\begin{equation}
  |(\mM_\Phi, f_a\otimes g_{-a})|\leq C_{\Phi,A'}\|f\|_{A,B}\|g\|_{A,B}
   e^{-2|\gamma\, a/A'|^2}
 \label{4.6}
\end{equation}
for each $A'>A$.

\medskip

 Proof.} We denote the vacuum expectation value~\eqref{2.3}
 by $M$ and set
 $\varphi=f\otimes g\otimes h_1\otimes h_2$,
 $\varphi_a=f_a\otimes g_{-a}\otimes h_1\otimes h_2$.  Then
\begin{equation}
  (\mM_\Phi, f_a\otimes g_{-a})=  (M, D_\theta \varphi_a).
 \label{4.7}
\end{equation}
Theorem~1 of~\cite{S2} shows that under the condition
$B<1/\sqrt{e|\theta|}$ the operator $D_\theta$ maps the space
$S^{1/2,B}_{1/2, A}$ continuously into the space $S^{1/2,B'}_{1/2,
A}$, where $B'=B\sqrt{2}$. In particular, $\|D_\theta
\varphi\|_{A,B'}\leq C\|\varphi\|_{A,B}$, which gives the
inequality
\begin{equation}
|\partial^\kappa (D_\theta\varphi)(x-a,y+a,z_1,z_2)|\leq
C\|\varphi\|_{A,B}B^{\prime
|\kappa|}\kappa^{\kappa/2}e^{-(|x-a|^2+
|y+a|^2+|z_1|^2+|z_2|^2)/A^2}.
 \label{4.8}
\end{equation}
Clearly, $M$ is a tempered distribution supported in the cone
 \begin{equation}
\bar \oV\times \oR^{3d}=\{(x,y,z_1,z_2)\in \oR^{4d}\colon
(x-y)^2\geq 0\}.
 \notag
\end{equation}
Therefore there exist an integer $N$ and a constant $C'$ such
that, for each test function $\psi\in \mS(\oR^{4d})$, we have
\begin{equation}
|(M, \psi)|\leq C'\|\psi\|_{N,\bar \oV\times \oR^{3d}},
 \label{4.9}
\end{equation}
where
\begin{equation}
\|\psi\|_{N,\bar \oV\times \oR^{3d}}=\sup_{|\kappa|\leq
N}\sup_{\bar \oV\times
\oR^{3d}}\left(1+|x|+|y|+|z_1|+|z_2|\right)^N
|\partial^\kappa\psi(x,y,z_1,z_2)|.
 \label{4.10}
\end{equation}
We put $\psi=D_\theta \varphi_a$ and denote $x-y$ by $\xi$.
Combining~\eqref{4.7}--\eqref{4.10}, we obtain
\begin{multline}
|(\mM_\Phi, f_a\otimes g_{-a})| \\
\leq C^{\prime\prime}\|\varphi\|_{A,B}\sup_{\bar \oV\times
\oR^{3d}}\left(1+|x|+|y|+|z_1|+|z_2|\right)^N
e^{-(|x-a|^2+ |y+a|^2+|z_1|^2+|z_2|^2)/A^2}\\
\leq C_{h_1,h_2}\|f\otimes g\|_{A,B}\sup_{\xi\in
\bar\oV}(1+|\xi|)^Ne^{-|\xi-2a|^2/(2A^2)}.
 \label{4.11}
\end{multline}
To complete the proof it suffices to observe that
$$
|\xi-2a|\geq 2\gamma|a|,\quad |\xi-2a|\geq \gamma|\xi|\quad \text
{for all}\quad \xi\in \bar\oV.
$$

The obtained estimate~\eqref{4.6}  is the stronger, the smaller
 $A$. However the space $S^{1/2,B}_{1/2,
A}$ becomes trivial if $AB$ is too small. For the readers'
convenience, a proof of this simple fact is given in the appendix.
If $AB>2/\sqrt{e}$, then $S^{1/2,B}_{1/2, A}$ is nontrivial and,
in particular, contains the Gaussian function $e^{-2|x/A|^2}$.
Because of the restriction $B<1/\sqrt{e|\theta|}$ in the
assumptions of Theorem~2, the best result is at  $A\sim
2\sqrt{|\theta|}$. It can be interpreted as demonstrating that the
matrix element~\eqref{4.5} decreases like $e^{-|\gamma
a|^2/(2|\theta|)}$ at large spacelike separation of the test
functions along the direction $a$, which refines the statement
made at the beginning of this section.

\section{\large The star commutator}

In Refs.~\cite{FW,Ch}, a framework for noncommutative QFTs was
formulated in terms of the vacuum expectation values of
$\star$-products of field operators at different spacetime points.
 This product is formally written as
\begin{equation}
\phi(x_1)\star\dots\star
\phi(x_n)=\prod_{a<b}e^{(i/2)\partial_{x_a}\theta\partial_{x_b}}
\phi(x_1)\cdots\phi(x_n).
 \label{5.1}
 \end{equation}
It is generally agreed that a mathematically rigorous theory of
quantum fields on noncommutative spacetime shall adopt the basic
assumption of the traditional axiomatic approach~\cite{SW,BLOT}
that quantum fields  are operator-valued  distributions. In other
words, it is customary to assume that in this case, too, there is
a linear mapping of the Schwartz space $\mS(\oR^d)$ (or another
suitable test function space) into the operators of the Hilbert
space of states: $f\longrightarrow \phi(f)$. This raises the
question of a rigorous definition of the formal
expression~\eqref{5.1} in agreement with this assumption. First of
all, we note that there is a multilinear mapping
$\underbrace{\mS(\oR^d)\times\cdots \times \mS(\oR^d)}_{n}\to
\mS(\oR^{nd})$ associated naturally with the Moyal $\star$-product
$(f_1,\dots, f_n)\to f_1\star\cdots\star f_n$. It is defined by
\begin{multline}
(f_1,\dots, f_n)\longrightarrow f_1(x_1)\star\cdots\star f_n(x_n)\\
=\frac{1}{(2\pi)^{dn}}\int dk_1\dots dk_n\tilde f_1(k_1)\cdots
\tilde f_n(k_n)e^{-i\sum_a k_a\cdot
x_a}\prod_{a<b}e^{-(i/2)k_{a\mu}\theta^{\mu\nu}k_{b\nu}}.
 \label{5.2}
 \end{multline}
The notation $f_1(x_1)\star\cdots\star f_n(x_n)$ is accepted in
the literature, though it seems reasonable to denote the
function~\eqref{5.2} by $f_1\otimes_\star\dots\otimes_\star f_n$.
The ordinary product of $n$ functions $f_1,\dots, f_n$ is obtained
from  $(f_1\otimes\dots\otimes f_n)(x_1,\dots,x_n)$
 by the identification $x_1=\dots=x_n$, and their  Moyal product
 is obtained from~\eqref{5.2} in the same fashion. Sometimes we
 will write $f_1\otimes_\star\dots\otimes_\star f_n$ instead of
 $f_1(x_1)\star\cdots\star f_n(x_n)$ to avoid confusion and for short.
If the test functions are sufficiently smooth, then~\eqref{5.2}
can be rewritten as
\begin{equation}
f_1(x_1)\star\cdots\star f_n(x_n)=\prod_{a<b}
e^{(i/2)\partial_{x_a}\theta\partial_{x_b}}f_1(x_1)\cdots
f_n(x_n).
 \label{5.3}
 \end{equation}
In particular, the power series expansion of the expression on the
right-hand side of~\eqref{5.3} in $\theta$ converges to the
function~\eqref{5.2} in the space $\mS^{1/2}(\oR^{dn})$ whose
elements satisfy the inequalities~\eqref{4.1} for each $A>0$ (with
a constant $C_\kappa$ depending on  $f$ and $A$). The topology of
$\mS^{1/2}$ is defined by the system of norms corresponding to
these inequalities. As shown in~\cite{S1,S2}, $\mS^{1/2}$ is the
largest subspace of the Schwartz space with such a convergence
property. The operation $(f_1,\dots, f_n)\to
f_1\otimes_\star\cdots\otimes_\star f_n$ generates a dual
operation over the distributions $u_j\in \mS'(\oR^d)$, which is
equivalent to multiplication of $\tilde u_1\otimes\dots\otimes
\tilde u_n$ by the multiplier
\begin{equation}
\mu_n=\prod_{1\leq a<b\leq
n}e^{-(i/2)k_{a\mu}\theta^{\mu\nu}k_{b\nu}}.
 \label{5.4}
 \end{equation}
 In particular, in the case of two distributions we have
\begin{equation}
(u\otimes_\star v)(x,y)\equiv u(x)\star
v(y)=\frac{1}{(2\pi)^{2d}}\int dkdq\, e^{-ik\cdot x-iq\cdot
y-i[k,q]}\tilde u(k)\tilde v(q),
 \label{5.5}
 \end{equation}
with the above notation $[k,q]= (1/2)k_\mu\theta^{\mu\nu}q_\nu$.
This operation over distributions can also be considered as an
extension of the operation~\eqref{5.3} over test functions by
continuity. The extension is unique, because $\mS^{1/2}$ is dense
in $\mS'$.

Now let $\phi$ be an operator-valued tempered distribution defined
on a dense invariant domain $D$ in the Hilbert space $\mH$, with
the vacuum vector $\Psi_0\in D$. By the standard
arguments~\cite{SW} based on the Schwartz kernel theorem, the
vector
\begin{equation}
\Phi_n (f)=\int dx_1\dots dx_n\, \phi(x_1)\cdots\phi(x_n)f(x_1,
\dots, x_n)\Psi_0
 \label{5.6}
 \end{equation}
 and the operator  $\int dx_1\dots dx_n\,
 \phi(x_1)\cdots\phi(x_n)f(x_1,\dots,x_n)$ are well-defined for each $f\in
\mS(\oR^{dn})$. In particular, the operator
\begin{equation}
\int dx_1\dots dx_n\,
\phi(x_1)\cdots\phi(x_n)f_1(x_1)\star\cdots\star f_n(x_n),
 \label{5.7}
 \end{equation}
is uniquely defined for any system of functions $f_a\in
\mS(\oR^d)$, $a=1,\dots n$. An analogous statement  holds in the
case when $\mS(\oR^d)$ is  replaced by another nuclear space which
is a topological algebra under the $\star$-product, for instance,
by the space $\mS^{1/2}(\oR^d)$. If we hold to the basic principle
of the calculus of generalized functions and define the action of
the differential operator in~\eqref{5.1} by duality, then
\begin{multline}
\int dx_1\dots
dx_n\,\phi(x_1)\star\cdots\star\phi(x_n)f_1(x_1)\cdots f_n(x_n)\\=
\int dx_1\dots
dx_n\,\phi(x_1)\cdots\phi(x_n)f_1(x_1)\star\dots\star f_n(x_n).
 \label{5.8}
 \end{multline}
As a consequence, we obtain the relation
\begin{multline}
\int dx_1\dots dx_n\, W_\star^{(n)}(x_1,\dots,x_n)f_1(x_1)\cdots f_n(x_n)\\
=\int dx_1\dots
dx_n\,W^{(n)}(x_1,\cdots,x_n)f_1(x_1)\star\dots\star f_n(x_n),
 \label{5.9}
\end{multline}
where $W^{(n)}(x_1,\dots, x_n)=\langle \Psi_0,
\phi(x_1)\cdots\phi(x_n)\Psi_0\rangle$ is the usual Wightman
function and
\begin{equation}
W_\star^{(n)}(x_1,\dots, x_n)\eqdef\langle \Psi_0,
\phi(x_1)\star\cdots\star\phi(x_n)\Psi_0\rangle.
 \label{5.10}
 \end{equation}
 We note that~\eqref{5.9} can also be written as
\begin{equation}
(W_\star^{(n)},f_1\otimes\cdots\otimes
f_n)=(W^{(n)},f_1\otimes_\star\cdots\otimes_\star f_n).
 \notag
\end{equation}
Clearly, the two-point function $W(x,y)$  coincides with
$W_\star(x,y)$ due to the translation invariance. Indeed, writing
$W(x,y)=w(x-y)$, we have
\begin{multline}
\int dxdy\,w(x-y)f(x)\star g(y)=
 \frac{1}{(2\pi)^d}\int dxdy dk dq\,
 \tilde w(k)\delta(k+q) e^{-ik\cdot x-iq\cdot y} f(x)\star g(y)
\\= \frac{1}{(2\pi)^d}\int dk dq\,\tilde w(k)\delta(k+q) \tilde f(-k)
\tilde g(-q)e^{-i[k, q]}=\int dxdy\,w(x-y)f(x)g(y),
 \notag
\end{multline}
because  $e^{-i[k, q]}=1$  for $k=-q$. But for $n>2$, the
distributions $W^{(n)}$ and $W_\star^{(n)}$ differ from one
another.

Let us consider the definition~\eqref{5.8} more closely, taking a
free field $\phi$ as a simplest example. Let $w(x-y)$ be its
two-point function. It is easy to see that if the product
$\phi(x)\star\phi(y)$ is defined by~\eqref{5.8}, then
\begin{equation}
 \langle 0|\phi(x)\star\phi(y):\phi(z_1)\phi(z_2):|0\rangle=
 w(x-z_1)\star w(y-z_2)+ w(x-z_2)\star w(y-z_1).
 \label{5.11}
\end{equation}
Indeed, by the Wick theorem we have
\begin{equation}
 \langle 0|\phi(x)\phi(y):\phi(z_1)\phi(z_2):|0\rangle=
 w(x-z_1) w(y-z_2)+ w(x-z_2) w(y-z_1).
 \label{5.12}
\end{equation}
Let $f$, $g$, $h_1$, $h_2$ be functions in the Schwartz space.
Using~\eqref{5.5}, \eqref{5.8},  and~\eqref{5.12}, we obtain
\begin{multline}
\int  dxdy dz_1dz_2\,\langle
0|\phi(x)\star\phi(y):\phi(z_1)\phi(z_2):|0\rangle
f(x)g(y)h_1(z_1)h_2(z_2)\\= \int dz_1dz_2\int  dxdy\,[w(x-z_1)
w(y-z_2)+ w(x-z_2) w(y-z_1)]f(x)\star g(y)h_1(z_1)h_2(z_2)\\= \int
dz_1dz_2\int \frac{dkdq}{(2\pi)^{2d}} \left[e^{ik\cdot z_1+iq\cdot
z_2}+e^{ik\cdot z_2+iq\cdot z_1}\right]\tilde w(k)\tilde w(q)
e^{-i[k, q]} \tilde f(-k)\tilde g(-q) h_1(z_1)h_2(z_2)\\= \int
dxdy dz_1dz_2\,[w(x-z_1)\star w(y-z_2)+ w(x-z_2)\star w(y-z_1)]
f(x)g(y)h_1(z_1)h_2(z_2),
 \label{5.13}
\end{multline}
which proves our claim. The formula~\eqref{5.11} is also
obtainable by applying formally the operator
$e^{(i/2)\partial_x\theta
\partial_y}$ to~\eqref{5.12}. In momentum space, the
distribution~\eqref{5.11} takes the form
\begin{equation}
(2\pi)^{2d}\tilde w(k)\tilde w(q) e^{-i[k,
q]}[\delta(k+p_1)\delta(q+p_2)+\delta(k+p_2)\delta(q+p_1)],
 \label{5.14}
\end{equation}
where the variables $k,q,p_1,p_2$ are,  respectively, conjugate to
the coordinate-space variables $x,y,z_1,z_2$. We note
that~\eqref{5.14} differs from the Fourier transform of the
distribution~\eqref{5.12}  only by the factor $e^{-i[k, q]}$.

In~\cite{Ch2, Ch}, it was assumed that in the case of space-space
noncommutativity, the star commutator
$[\phi(x),\phi(y)]_*=\phi(x)\star\phi(y)-\phi(y)\star\phi(x)$
obeys microcausality with respect to the commuting coordinates
$(x^0, x^3)$, i.e.,  $[\phi(x),\phi(y)]_*=0$ everywhere in the
wedge $\{(x,y)\in \oR^{2d}\colon |x^0-y^0|<|x^3-y^3|\}$. We shall
show that this assumption contradicts  the spectral condition if
the product~\eqref{5.1} is defined by duality, as indicated above.

\medskip

{\bf Theorem 3.} {\it Let $\phi$ be a free neutral scalar field on
$\oR^d$ and let $\Phi$ be a two-particle state  of the
form~\eqref{3.2}. If the product $\phi(x)\star\phi(y)$ is defined
by~\eqref{5.8}, then the distribution
\begin{equation}
 \langle 0|[\phi(x),\phi(y)]_\star|\Phi\rangle
 \label{5.15}
\end{equation}
does not vanish on any  open set and so its support coincides with
the whole space $\oR^{2d}$.

\medskip

Proof.} From~\eqref{5.14}, it follows that the Fourier transform
of $\langle 0|\phi(x)\star\phi(y)|\Phi\rangle$ is of the form
\begin{equation}
2\tilde w(k)\tilde w(q) e^{-i[k, q]}\tilde h(k)\tilde h(q).
 \label{5.16}
 \end{equation}
The Fourier transform of  $\langle
0|\phi(y)\star\phi(x)|\Phi\rangle$ is obtained from~\eqref{5.16}
by interchanging $k$ and $q$. Hence that of the matrix
element~\eqref{5.15} has the form
\begin{equation}
-4 i \tilde w(k)\tilde w(q) \sin[k, q]\tilde h(k) \tilde h(q)
 \notag
\end{equation}
and differs from $\tilde w\otimes\tilde w$ only by the factor
$-4i\sin[k,q]\tilde h(k)\tilde h(q)$ which is a multiplier of the
Schwartz space and does not vanish on  $\supp (\tilde
w\otimes\tilde w)$ if $\Phi\ne 0$. The support of $\tilde
w\otimes\tilde w$ is contained in the properly convex cone
$\oV_+\times\oV_+$. Therefore, the distribution~\eqref{5.15} is
the boundary value of a function analytic in the tubular domain
$\oR^{2d}+i(\oV_-\times \oV_-)$ (see, e.g.,~\cite{BLOT},
Theorem~B.7). Applying the generalized uniqueness theorem (ibid,
Theorem~B.10), we conclude that  this distribution does not vanish
on any nonempty open set, because otherwise it would be
identically zero on $\oR^{2d}$. Theorem~3 is proved.

Now we return to the sample observable $\mathcal O(x)$ defined
by~\eqref{2.4} and consider the star commutator
\begin{equation}
 [\mathcal
O(x),\mathcal O(y)]_\star= \mathcal O(x)\star\mathcal
O(y)-\mathcal O(y)\star\mathcal O(x).
 \label{5.17}
\end{equation}

\medskip

{\bf Theorem 4.} {\it Let, as in Theorem~1, $d=4$,
$\theta^{12}=-\theta^{21}\ne 0$, and the other elements of the
matrix $\theta^{\mu\nu}$ be equal to zero. Then the star
commutator~\eqref{5.17} does not vanish in the wedge defined
by~\eqref{2.11}.

\medskip

Proof.} Let  $(\bar x, \bar y)$ be contained in the
wedge~\eqref{2.11} together with a neighborhood $U\times V$. In
what follows we set $U=U_c\times U_{nc}$, $V=V_c\times V_{nc}$,
where the labels $c$ and $nc$ indicate,  respectively, sets in the
planes $(x^0,x^3)$ and $(x^1,x^2)$. For definiteness, we assume
that $|\bar x^0- \bar y^0|< \bar x^3- \bar y^3$ and
\begin{equation}
U_c-V_c\subset \oV_R,
 \label{5.18}
\end{equation}
where $\oV_R$ is the right component of the spacelike cone  in
$\oR^2$. We shall show that there exist test functions $f,g$
supported in $U$, $V$ and a state $\Phi$ of the form~\eqref{3.2}
such that the matrix element
\begin{equation}
 \langle 0|\int  dxdy\,[\mathcal
O(x),\mathcal O(y)]_\star f(x)g(y)|\Phi\rangle \label{5.19}
\end{equation}
is different from zero. Applying the operator
$e^{(i/2)\partial_x\theta
\partial_y}$ to~\eqref{2.7}, we
obtain
\begin{multline}
 \langle 0|\,\mathcal O(x)\star\mathcal
O(y):\phi(z_1)\phi(z_2)\!:|0\rangle=
 4\!\int\! dkdp_1dp_2\,\tilde w(k)
  e^{-ik\cdot (x-y)-ip_1\cdot (x-z_1)-ip_2\cdot (y-z_2)}\\
  \times  e^{-i[k,p_1+p_2]-i[p_1,p_2]}
\prod_{i=1,2}\tilde w(p_i)\cos[k,p_i] + (z_1\leftrightarrow z_2).
 \label{5.20}
 \end{multline}
On the other hand, applying $e^{(i/2)\partial_y\theta
\partial_x}$ to $\mW(y,x;z_1,z_2)$ gives
\begin{multline}
 \langle 0|\,\mathcal O(y)\star\mathcal
O(x):\phi(z_1)\phi(z_2)\!:|0\rangle=
 4\!\int\! dkdp_1dp_2\,\tilde w(k)
   e^{-ik\cdot (y-x)-ip_1\cdot (y-z_1)-ip_2\cdot
(x-z_2)}\\
\times e^{-i[k,p_1+p_2]-i[p_1,p_2]}\prod_{i=1,2}\tilde
w(p_i)\cos[k,p_i] + (z_1\leftrightarrow z_2).
 \label{5.21}
 \end{multline}
From~\eqref{5.20} and \eqref{5.21}, it follows that
\begin{multline}
 \langle 0|\,[\mathcal O(x),\mathcal
O(y)]_\star :\phi(z_1)\phi(z_2)\!:|0\rangle=
 8i\int\! dkdp_1dp_2\,\tilde \Delta(k)
  e^{-ik\cdot (x-y)-ip_1\cdot (x-z_1)-ip_2\cdot
(y-z_2)}\\
\times \cos([k,p_1+p_2]+[p_1,p_2])\prod_{i=1,2}\tilde w(p_i)\cos[k,p_i]\\
-8i\int\! dkdp_1dp_2\,\tilde \Delta_1(k) e^{-ik\cdot
(x-y)-ip_1\cdot (x-z_1)-ip_2\cdot
(y-z_2)}\\
\times \sin([k,p_1+p_2]+[p_1,p_2])\prod_{i=1,2}\tilde
w(p_i)\cos[k,p_i] + (z_1\leftrightarrow z_2),
 \label{5.22}
 \end{multline}
where $\tilde \Delta_1(k)=\tilde w(k)+\tilde
w(-k)=2\pi\delta(k^2-m^2)$. The distribution defined by the first
integral on the right-hand side of~\eqref{5.22} and that obtained
from it by the transposition $z_1\leftrightarrow z_2$ vanish in
the wedge~\eqref{2.11} by  the argument that was used at the end
of Sec.~2. Now we consider the distribution
\begin{multline}
 \mathcal W(x-y,x-z_1, y-z_2)=
 -8i\int\! dkdp_1dp_2\,\tilde w(k)
    e^{-ik\cdot (x-y)-ip_1\cdot (x-z_1)-ip_2\cdot
(y-z_2)}\\
\times \sin([k,p_1+p_2]+[p_1,p_2])\prod_{i=1,2}\tilde
w(p_i)\cos[k,p_i].
  \label{5.23}
 \end{multline}
Clearly, it is not identically zero because there are points $k,
p\in \supp \tilde w$ such that $\cos[k,p]\ne 0$ and $\sin[k,2p]\ne
0$. Moreover, there exists a function $h\in \mS(\oR^4)$ such that
the distribution
\begin{equation}
T(x,y)=\int dz_1dz_2\, \mathcal W(x-y,x-z_1, y-z_2) h(z_1)h(z_2)
\label{5.24}
\end{equation}
is also nonzero. The spectral condition $\supp\tilde w\subset \bar
\oV_+$ implies that $T$ is the boundary value of a function
analytic in the tubular domain defined by
\begin{equation}
\Im(x-y)\in \oV_-,\qquad \Im x\in \oV_-,\qquad \Im y\in \oV_-.
\label{5.25}
\end{equation}
By the generalized uniqueness theorem, $T$ does not vanish on any
open subset of $\oR^{8}$ and, in particular, on $U\times V$. Every
test function supported in $U\times V$ can be approximated by
linear combinations of functions of the form $f\otimes g$, where
$\supp f\subset U$, $\supp g\subset V$. Therefore, there  are
$f,g$ supported in these neighborhoods and such that $(T,f\otimes
g)\ne0$. The matrix element~\eqref{5.19} is written
\begin{equation}
\int dxdydz_1dz_2\,(\mathcal W(x-y,x-z_1, y-z_2)+\mathcal
W(y-x,x-z_1, y-z_2)) f(x)g(y)h(z_1)h(z_2)
  \label{5.26}
 \end{equation}
and, to complete the proof, it suffices to show that the
expression~\eqref{5.26} is equal to $2(T, f\otimes g)$. Clearly,
we can assume that each of the functions $f$, $g$, $h$ is the
product of functions of the commuting coordinates and of the
noncommuting coordinates which are, respectively, labelled $c$ and
$nc$ below. Let
\begin{multline}
 \mathcal W_c(x^c-y^c,x^c-z^c_1, y^c-z^c_2)\eqdef
 \int\! dx^{nc}dy^{nc}dz^{nc}_1dz^{nc}_2\,
 \mathcal W(x-y,x-z_1, y-z_2)\\
\times f_{nc}(x^{nc})
g_{nc}(y^{nc})h_{nc}(z_1^{nc})h_{nc}(z^{nc}_2),
  \label{5.27}
 \end{multline}
and let $\xi=x^c-y^c$, $\zeta_1= x^c-z^c_1$, $\zeta_2=y^c-z^c_2$.
If $\chi\in \mS(\oR^4)$, then by the spectral condition the
distribution $T_\chi(\xi)=\int \mathcal W_c(\xi, \zeta_1,
\zeta_2)\chi(\zeta_1,\zeta_2)d\zeta_1 d\zeta_2$ is the boundary
value of a function analytic in the tubular domain whose base is
the lower cone  $\{\xi\in \oR^2\colon \xi^2>0, \xi^0<0\}$. This
analytic function is invariant under Lorentz boosts and, by  the
simplest two-dimensional version of the Bargman-Hall-Wightman
theorem~\cite{SW,BLOT}, admits  analytic continuation to an
extended domain which contains all spacelike points of $\oR^2$.
The analytic extension is invariant under the reflection $\xi\to
-\xi$ and hence $\int d\xi\,T_\chi(\xi)\psi(\xi)=\int d\xi\,
T_\chi(-\xi)\psi(\xi)$ for every $\psi\in \mS(\oR^2)$ whose
support is contained in ${\oV}_R$. It follows that
$$
\int d\xi d\zeta_1 d\zeta_2\, \mathcal W_c(\xi, \zeta_1,
\zeta_2)\varphi(\xi,\zeta_1\zeta_2)= \int d\xi d\zeta_1 d\zeta_2\,
\mathcal W_c(-\xi, \zeta_1, \zeta_2)\varphi(\xi,\zeta_1\zeta_2)
$$
for every $\varphi\in\mS(\oR^6)$ whose support is contained in
${\oV}_R\times\oR^2\times\oR^2$. The function
\begin{equation}
\varphi(\xi, \zeta_1, \zeta_2)=\int
f_c(X)g_c(X-\xi)h_c(X-\zeta_1)h_c(X-\xi-\zeta_2)dX
 \label{5.28}
 \end{equation}
 has support in this wedge by construction and, for this function,
 we have
 $$
(\mathcal W_c,\varphi)=\int dxdydz_1dz_2\,\mathcal W(x-y,x-z_1,
y-z_2) f(x)g(y)h(z_1)h(z_2) .
$$
Thus the expression~\eqref{5.26} is indeed equal to $2(T, f\otimes
g)$. Theorem~4 is proved.

\section{\large Concluding remarks}

At the present time, there is no agreement regarding the physical
interpretation of the $\star$-product of quantum fields at
different spacetime points. In this connection we shall make some
remarks about the proposals to formulate a framework for quantum
field theories on noncommutative spacetime in terms of the
$n$-point vacuum expectation values of such products (see, e.g.,
Refs.~\cite{FW, Ch}). Let $\phi$ be a scalar field with test
functions in $\mS(\oR^d)$ and with an invariant dense domain $D$
in the Hilbert space $\mH$, containing the vacuum state $\Psi_0$.
If the star-modified Wightman functions $W_\star^{(n)}$ of $\phi$
are defined by~\eqref{5.10}, \eqref{5.8}, then we always can
construct a field $\phi_\theta$ such that
\begin{equation}
\langle
\Psi_0,\phi_\theta(x_1)\phi_\theta(x_2)\cdots\phi_\theta(x_n)\Psi_0\rangle=
W_\star^{(n)}(x_1,x_2,\dots, x_n). \label{6.1}
 \end{equation}
Indeed, let $g\in \mS(\oR^d)$, $f\in \mS(\oR^{dn})$, and let
$\Phi_n(f)$ be a vector of the form~\eqref{5.6}. We set
\begin{equation}
\phi_\theta(g)\Psi_0=\phi(g)\Psi_0,\qquad
\phi_\theta(g)\Phi_n(f)=\Phi_{n+1}(g\otimes_\star f),\quad n\geq
1,\label{6.2}
\end{equation}
where
\begin{equation}
 (g\otimes_\star f)(y,x_1,\dots, x_n)\eqdef\prod_{a=1}^n
e^{(i/2)\partial_y\theta\partial_{x_a}}g(y)f(x_1,\dots x_n).
 \label{6.3}
 \end{equation}
Then~\eqref{6.1} is satisfied. Clearly,  $W^{(n)}_\star\in
S'(\oR^{dn})$ and, for each $\tilde f\in S(\oR^{dn})$,
$$
(\tilde W^{(n)}_\star,\tilde f)=(\tilde W^{(n)},\mu_n\cdot \tilde
f),
$$
where $W^{(n)}$ is the ordinary $n$-point Wightman function of
$\phi$ and $\mu_n$ is given by~\eqref{5.4}. The distributions
$W_\star^{(n)}$, i.e., the vacuum expectation values of the
ordinary products of the  operators $\phi_\theta$, have the same
spectral properties as $W^{(n)}$ because the multiplication by
$\mu_n$ leaves these properties unchanged. If the field $\phi$ is
Hermitian, then so are $\phi_\theta$. Indeed, for any $g\in
\mS(\oR^d)$, $f\in \mS(\oR^{dn})$, $h\in \mS(\oR^{dm})$, we have
\begin{equation}
\langle\Phi_m(h),\Phi_{n+1}(g\otimes_\star f)\rangle=
\langle\Phi_{m+1}(\bar g\otimes_\star h),\Phi_n(f)\rangle,
 \label{6.4}
 \end{equation}
where the bar over $g$ denotes the complex conjugation. This
identity is easily  verified by using the antisymmetry of
$\theta^{\mu\nu}$ and going to the momentum-space representation
because $\tilde W^{(m+n+1)}(q_1,\dots, q_m, k, p_1,\dots,p_n)$
contains the factor $\delta(k+\sum_1^m q_b+\sum_1^n p_a)$ due to
the translation invariance. For the same reason the modified
Wightman functions satisfy the ordinary positive definiteness
conditions
\begin{equation}
\sum_{m,n=1}^N (W^{(m+n)}_\star,f^\dagger_m\otimes f_n)\geq 0,
 \label{6.5}
 \end{equation}
where $f_n$ are arbitrary elements of  $\mS(\oR^{dn})$ and
$f^\dagger(x_1,\dots, x_n)\eqdef \overline{f(x_n,\dots, x_1)}$.
 To prove~\eqref{6.5}, it is enough to observe that
 $(W^{(m+n)},g\otimes f)=(W^{(m+n)},g\otimes_\star f)$
for any  $g\in \mS(\oR^{dm})$ and $f\in \mS(\oR^{dn})$, where
$g\otimes_\star f$ is defined by
\begin{equation}
(g\otimes_\star f)(y_1,\dots,y_m,x_1,\dots, x_n)=
\prod_{b=1}^m\prod_{a=1}^n
e^{(i/2)\partial_{y_b}\theta\partial_{x_a}}g(y_1,\dots,
y_m)f(x_1,\dots x_n). \label{6.6}
 \end{equation}
 However the transformation and local properties of $\phi_\theta$
 differ radically from those of    $\phi$.

For the case of a free field $\phi$, an alternate description of
its associated field $\phi_\theta$ is given in~\cite{GL}. Namely,
the creation and annihilation operators of these fields are
related by
$$
a_\theta(p)=e^{(i/2)p\,\theta P} a(p),\qquad
a^\dagger_\theta(p)=e^{-(i/2)p\,\theta P} a^\dagger(p),
$$
where  $P$ is the energy-momentum operator. The operators
$a_\theta(p)$, $a^\dagger_\theta(p)$ satisfy the deformed
commutation relations
\begin{gather}
a_\theta(p)a_\theta(p')=e^{-ip\theta
p'}a_\theta(p')a_\theta(p),\quad
a^\dagger_\theta(p)a^\dagger_\theta(p')=e^{-ip\theta
p'}a^\dagger_\theta(p')a^\dagger_\theta(p),\notag\\
a_\theta(p)a^\dagger_\theta(p')=e^{ip\theta
p'}a^\dagger_\theta(p') a_\theta(p)+2\omega_{\bf p}\delta(\bf p-
\bf p').
 \notag
 \end{gather}
As discussed above, the field $\phi_\theta$ is essentially
nonlocal, but fields with different $\theta$ have interesting
relative localization properties found by Grosse and
Lechner~\cite{GL}. On the other hand, Fiore and Wess~\cite{FW}
argued that  the twisted Poincar\'e covariance can be implemented
in the theory of a free field on noncommutative spacetime in a
manner  compatible with microcausality only if the canonical
commutation relations (CCR) of creation and annihilation operators
are suitably deformed. This deformation compensates the spacetime
noncommutativity and the vacuum expectation values of the
$\star$-products  defined in~\cite{FW} coincide with the usual
Wightman functions of a free field on commutative spacetime. In
other words, we obtain a mathematically self-consistent
formulation, but without a new physics and in fact even without
noncommutativity. The same disappointing conclusion has been drawn
in~\cite{FW} for interacting fields treated perturbatively. The
axiomatic scheme proposed for the star-modified Wightman functions
in~\cite{Ch} differs from that of~\cite{FW} and does not include a
deformation of the CCR algebra. But then the spectral condition
comes into conflict with causality as is evident from the
foregoing.

It is quite possible that microcausality should be replaced by a
weaker condition in order to develop a satisfactory framework for
quantum field theory on noncommutative spacetime. We believe that
the above-stated $\theta$-locality condition is a possible
candidate for this role because it precisely describes the
nonlocal character of  the Moyal $\star$-product. Conditions of
this kind were previously used in nonlocal QFT and, together with
the relativistic covariance and the  spectral condition, they
ensure the existence of CPT symmetry as well as the usual
spin-statistics relation for nonlocal fields, see~\cite{S99}. An
extension of these results to the noncommutative setting is
discussed in~\cite{S06}, where analogous theorems are proved for
the case of a charge scalar field and space-space noncommutativity
with the residual $SO_0(1,1)\times SO(2)$-symmetry.

\section* {\large Acknowledgments}

 This paper was supported in part by the the Russian
Foundation for Basic Research (Grant No.~05-01-01049) and the
Program for Supporting Leading Scientific Schools (Grant
No.~LSS-1615.2008.2).

\section*{\large Appendix}

{\bf Lemma.} {\it If $AB> 2/\sqrt{e}$, then the space
$S^{1/2,B}_{1/2, A}$ is nontrivial. If $AB< \sqrt{2/e}$, then it
contains only the trivial function which is identically zero.

\medskip

Proof.} If $f\in S^{1/2,B}_{1/2, A}$, then the function
$f_\lambda(x)=f(\lambda x)$, where $\lambda>0$, belongs to the
space $S^{1/2,\lambda B}_{1/2, A/\lambda}$. Therefore, if a pair
$(A_0,B_0)$ is admissible, i.e., defines a nontrivial space, then
all the pairs $(A,B)$ for which $AB=A_0B_0$ are also admissible.
Let us show that $e^{-|x|^2}$ belongs to any space
$S^{1/2,B}_{1/2, A}$ with $A=\sqrt{2}$ and  $B>\sqrt{2/e}$.
Because $e^{-|x|^2}=\prod_j e^{-x_j^2}$, it suffices to consider
the  one variable case. By  the Cauchy inequality,
$$
 \left|\partial^\kappa e^{-x^2}\right|\le \frac{\kappa!}{r^\kappa}
 \max_{|\zeta-x|=r}e^{-\Re \zeta^2}
 \eqno{(\rm A1)}
$$
 for each $r>0$. Setting $\zeta=x+r e^{i\alpha}$  and using the
 elementary relation $\cos2\alpha=2\cos^2\alpha-1$, we get
$$
\Re \zeta^2=x^2+2xr\cos\alpha+r^2\cos2\alpha\geq
\frac{x^2}{2}-r^2.
$$
Therefore,
$$
 \left|\partial^\kappa e^{-x^2}\right|\leq \kappa!\,e^{-x^2/2}\,
 \inf_{r>0} \frac{e^{r^2}}{r^\kappa}=\kappa!\,e^{-x^2/2}
 \left(\frac{2e}{\kappa}\right)^{\kappa/2}.
 \eqno{(\rm A2)}
$$
By the Stirling formula, we have $\kappa! \leq C_\epsilon
(1/e+\epsilon)^\kappa \kappa^\kappa$ for any $\epsilon>0$. The
first statement of Lemma is thus proved.

Now we prove the second statement. Here again,  it is suffices to
consider the one variable case.  From the
inequalities~\eqref{4.3}, it follows that every  $f\in
S^{1/2,B}_{1/2, A}(\oR)$ is an entire analytic function and hence
$$
f(\xi)= \sum_\kappa\frac{1}{\kappa!}(\xi-x)^\kappa \partial^\kappa
f(x)
 \eqno{(\rm A3)}
$$
for any $x,\xi\in \oR$.  Using~\eqref{4.3},  (\rm A3)  and
choosing
 $\bar B> B$, we  estimate $f(\xi)$ in the following way
$$
|f(\xi)|\leq Ce^{-|x/A|^2}\sum_\kappa \frac{1}{\kappa!}B^\kappa
\kappa^{\kappa/2}|\xi-x|^\kappa \leq C'e^{-|x/A|^2}\sup_\kappa
 \frac{1}{\kappa!}{\bar B}^\kappa
\kappa^{\kappa/2}|\xi-x|^\kappa,
 \eqno{(\rm A4)}
 $$
where $C'=C\,\sum_\kappa (B/{\bar B})^\kappa<\infty$. Using the
inequality $\kappa!\geq (\kappa/e)^\kappa$, we can replace the
upper bound in (\rm A4) by the function $M(e{\bar B}|\xi-x|)$,
where
$$
M(r)=\sup_{\kappa>0}
\frac{r^\kappa}{\kappa^{\kappa/2}}=e^{r^2/(2e)}.
 $$
 Because the point $x$ can be taken arbitrarily, (\rm A4) implies
 that $f(\xi)\equiv 0$ if $1/A^2> e{\bar B}^2/2$. This
completes the proof.

\end{document}